\documentclass[twocolumn,trackchanges]{aastex63}
\RequirePackage{lineno}
\usepackage{amsmath}
\usepackage{amssymb}
\usepackage{amsthm}
\usepackage{natbib,aasdefs,url,bm}
\usepackage{array}
\usepackage{float}
\usepackage{graphicx}
\usepackage{subfigure}   
\usepackage{color}
\usepackage{threeparttable} 
\usepackage{CJK}
\usepackage{lineno}
\usepackage{tablefootnote}
\usepackage{threeparttable}

\newcommand{\cy}[1]{\textcolor{black}{#1}}

\shorttitle{Revisiting X-ray Afterglows of Jetted TDEs with the External Reverse Shock}
\shortauthors{Yuan et al.}

\begin{document}
\begin{CJK*}{UTF8}{gbsn}

\title{Revisiting X-ray Afterglows of Jetted Tidal Disruption Events with the External Reverse Shock}

\correspondingauthor{Chengchao Yuan}
\email{chengchao.yuan@desy.de}

\author[0000-0003-0327-6136]{Chengchao Yuan (袁成超)}\affil{Deutsches Elektronen-Synchrotron DESY, 
Platanenallee 6, 15738 Zeuthen, Germany}

\author[0000-0001-7062-0289]{Walter Winter}\affil{Deutsches Elektronen-Synchrotron DESY, 
Platanenallee 6, 15738 Zeuthen, Germany}

\author[0000-0003-2478-333X]{B. Theodore Zhang (张兵)}\affil{Key Laboratory of Particle Astrophysics, Institute of High Energy Physics, Chinese Academy of Sciences, Beijing 100049, China}\affil{TIANFU Cosmic Ray Research Center, Chengdu, China}

\author[0000-0002-5358-5642]{Kohta Murase}
\affil{Department of Physics, Department of Astronomy \& Astrophysics, Center for Multimessenger Astrophysics, Institute for Gravitation and the Cosmos, The Pennsylvania State University, University Park, PA 16802, USA}\affil{Center for Gravitational Physics and Quantum Information, Yukawa Institute for Theoretical Physics, Kyoto University, Kyoto, Kyoto 606-8502, Japan}

\author[0000-0002-9725-2524]{Bing Zhang (张冰)}\affil{The Nevada Center for Astrophysics, University of Nevada, Las Vegas, Las Vegas, NV 89154, USA}\affil{Department of Physics and Astronomy, University of Nevada, Las Vegas, Las Vegas, NV 89154, USA}

\begin{abstract}
We investigate the {external reverse shock region} of relativistic jets as the origin of X-ray afterglows of jetted tidal disruption events (TDEs) that exhibit luminous jets accompanied by fast-declining non-thermal X-ray emissions. We model the dynamics of jet propagating within an external density medium, accounting for continuous energy injection driven by accretion activities. We compute the time-dependent synchrotron and inverse Compton emissions from the reverse shock region. Our analysis demonstrates that the reverse shock scenario can potentially explain the X-ray light curves and spectra of four jetted TDEs, AT 2022cmc, Swift J1644, Swift J2058, and Swift J1112. Notably, the rapid steepening of the late-stage X-ray light curves can be attributed jointly to the jet break and cessation of the central engine as the accretion rate drops below the Eddington limit. Using parameters obtained from X-ray data fitting, we also discuss the prospects for $\gamma$-ray and neutrino detection.  
\end{abstract}

\section{Introduction}\label{sec:intro}
Tidal disruption events (TDEs) occur when a star is torn apart by the tidal forces of a supermassive black hole \citep[SMBH, e.g.,][]{1975Natur.254..295H, 1988Natur.333..523R, 1989ApJ...346L..13E}, resulting in a transient lasting from months to years, visible across the electromagnetic spectrum, from the radio, infrared, optical/ultraviolet, to X-ray ranges \citep[e.g.,][]{2020SSRv..216...85S,2020SSRv..216...81A,2021ApJ...908....4V}.
A small fraction of TDEs exhibits luminous relativistic jet signatures. 
Since the discovery of the hard X-ray transient event J164449.3$+$573451 \citep[hereafter, Sw J1644,][]{2011Sci...333..203B,burrows2011relativistic,2011Sci...333..199L,2011Natur.476..425Z} by the Swift satellite, three additional jetted TDEs -- Swift J2058.4+0516 \citep[hereafter, Sw J2058,][]{2012ApJ...753...77C,2015ApJ...805...68P}, Swift J1112.2-8238 \citep[hereafter Sw J1112,][]{2015MNRAS.452.4297B,2017MNRAS.472.4469B}, and AT 2022cmc \citep{2022Natur.612..430A,2023NatAs...7...88P} -- have been recorded. The multi-wavelength observations of these four jetted TDEs have provided valuable prototypes for studying the radiation mechanisms, accretion histories, and jet dynamics over a time window of months to years \citep[e.g.,][]{2011MNRAS.416.2102G,2020NewAR..8901538D,2021SSRv..217...12D,2024arXiv240611513Y}. 

The four jetted TDEs exhibit prominent similarities in their X-ray and radio afterglows. The X-ray light curves can be roughly described by a power-law decay, with the index roughly ranging from $5/3$ \citep[e.g., Sw J1644,][]{2016ApJ...817..103M} to $2.2$ \citep[e.g., Sw J2058 and AT 2022cmc,][]{2024arXiv240410036E}. Late-time follow-up observations have revealed additional steepening in the X-ray light curves, suggesting common changes in jet evolution or the central engine \citep{2013ApJ...767..152Z,2015ApJ...805...68P,2024arXiv240410036E}. Additionally, the rapid variability on time scales of $\sim{\rm few}\times100-1000$ seconds imposes further constraints on the SMBH mass \citep{2015MNRAS.452.4297B,2016ApJ...817..103M,2023NatAs...7...88P,2024ApJ...965...39Y}, e.g., $\lesssim{\rm few}\times10^7~M_\odot$. In the radio and sub-millimeter bands, observations have shown that these emissions are typically long-lasting compared to the fast-declining X-ray emission. The radio/sub-millimeter light curves are consistent with synchrotron radiation from electrons accelerated by the {external forward shock of the jet} with a Lorentz factor of $\Gamma\sim1-10$ propagating through a circumnuclear medium \citep[e.g.,][]{2011Natur.476..425Z,2011MNRAS.416.2102G,2012MNRAS.420.3528M,2012ApJ...748...36B,2013ApJ...767..152Z,2016MNRAS.461.3375Y,2018ApJ...854...86E,2021ApJ...908..125C,2023MNRAS.522.4028M, 2024ApJ...963...66Z, 2024arXiv240611513Y}. However, the origin of jetted TDE X-ray afterglows still remains unclear.

The X-ray emission from jetted TDEs is likely produced in a separate emission region. Theoretical models involving jets powered by SMBH spin energy via large-scale magnetic fields \citep{2014MNRAS.437.2744T,2014MNRAS.445.3919K}, energy dissipation within magnetically dominated jets \citep{burrows2011relativistic}, variable accretion near the SMBH horizon \citep{2012Sci...337..949R}, jet shell collisions \citep{2013MNRAS.434.3463Z} and internal dissipations \citep{2012MNRAS.421..908W,2013ApJ...762...98L,Lei:2015fbf}, inverse Compton scattering of external photons \citep{2011Sci...333..203B,2016MNRAS.460..396C}, \cy{and synchrotron emission \citep{2024ApJ...965...39Y}} have been investigated to explain the X-ray observations of jetted TDEs. The power-law decaying X-ray light curves align with the mass fallback rates from complete \citep[$t^{-5/3}$, e.g.,][]{1988Natur.333..523R,1989IAUS..136..543P} and partial \citep[$t^{-2.2}$, e.g.,][]{Guillochon:2012uc} disruptions. The late-stage sharp steepening of the X-ray light curves may result from jet shutoff, caused by the accretion disk transitioning from thick to thin as the accretion rate decreases from super- to sub-Eddington states \citep{2013ApJ...767..152Z,2014MNRAS.437.2744T}. \cite{2014ApJ...784...87S} suggests that the accretion rate may undergo dramatic drop at that time the accretion disk becomes radiatively cooling and gas-pressure dominated. 

Moreover, two-component jet model with a fast inner component and slow outer component, has also been exploited to explain the multi-wavelength emission from TDEs \citep{2014ApJ...788...32W,2015MNRAS.450.2824M,2015ApJ...798...13L,2023ApJ...957L...9T,2024arXiv240413326S,2024arXiv240611513Y}. 
{Recently, \cite{2024arXiv240611513Y} suggested that the external reverse shock powered by a relativitisc jet launched from the active central engines, such as continuous energy injection associated with accretion, can also explain the X-ray observations of AT 2022cmc, where the late-time rapid decay can be attributed to jet breaks. This is also analogous to the long-lasting reverse shock model for shallow-decay afterglow emission of gamma-ray bursts (GRBs).}

Following \cite{2024arXiv240611513Y}, we investigate the external reverse shock scenarios to explain the X-ray spectra and light curves of the four jetted TDEs, \cy{focusing on the fast jet scenario with $\Gamma>10$, whereas a slow jet (e.g., $\Gamma\lesssim5$) may be responsible for the radio emission}. We present a generic, self-consistent model based on the TDE accretion history and multi-wavelength observations to describe the jet evolution and the time-dependent emissions in jet reverse shock regions. The motivation is that jet deceleration, combined with an active central engine, jointly determines the reverse shock emission, which can naturally reproduce the $t^{-\delta}~(\delta\sim5/3-2.2)$ X-ray afterglows. Additionally, the cessation of power injection would result in sharply decaying reverse shock emission. 

We model the time-dependent accretion rate, jet evolution within an external medium incorporating continuous power injection, and reverse shock emission in Sec. \ref{sec:model}. In Sec. \ref{sec:results}, we apply the reverse shock model to fit the X-ray light curves and spectra of four jetted TDEs and discuss the $\gamma$-ray and neutrino detectabilities. We discuss and conclude our work in Sec. \ref{sec:discussion} and Sec. \ref{sec:conclusion}. 

Throughout the paper, we use $T_{\rm obs}$, $t$, and $t'$ to denote the times measured in the observer's frame, the SMBH-rest frame, and the jet comoving frame, respectively. 

\section{TDE jet and reverse shock modeling}\label{sec:model}

\subsection{Accretion history}
Considering a TDE originated from the disruption of a main-sequence star of mass $M_\star$ by a SMBH of mass $M_{\rm BH}$, we write down the critical tidal radius $R_{\rm T}=f_{\rm T}(M_{\rm BH}/M_\star)^{1/3}R_\star$ \citep[e.g.,][]{1988Natur.333..523R}, where $f_{\rm T}\sim 0.02-0.3$ represents the correction from the stellar structures \citep[e.g.,][]{1989IAUS..136..543P,2015ApJ...806..164P} and $R_\star$ is the radius of the star. 
After the disruption, approximately half of the star's mass remains bound within an eccentric orbit. A fraction ($\eta_{\rm acc}$) of the bounded mass will end up being accreted by the SMBH. The fall back time can be estimated as the orbital period of most tightly bound matter, e.g., \cy{$t_{\rm fb}\approx2\pi\sqrt{a_{\rm min}^3/GM_{\rm BH}}\simeq3.3\times10^6{~\rm s}~f_{\rm T,-1.2}^{1/2}M_{\rm BH,7}^{1/2}M_{\star,0.7}^{-1/10}$ \citep[e.g.,][]{2020ApJ...902..108M}, where $ a_{\rm min}\approx R_{\rm T}^2/(2R_\star)$ is the semimajor axis of the orbit} and the stellar mass-radius relationship $R_{\star}=R_{\odot}(M_\star/M_\odot)^{1-\xi}$ \citep{1990sse..book.....K} with $\xi\sim0.4$ in the mass range $1<M_\star/M_\odot<10$ is adopted. \cy{Here, we chose $f_{\rm T}=10^{-1.2}$ as the fiducial value and scale $t_{\rm fb}$ to the value for AT 2022cmc, $M_\star=5M_{\star,0.7}M_{\odot}$ \citep{2024arXiv240611513Y}, for illustration purposes.}

Using $t_{\rm fb}$ and $\eta_{\rm acc}$, and considering that the post-fallback accretion rate follows a power-law of $t^{-5/3}$, we express the time-dependent accretion rate as \citep{2024arXiv240611513Y}
\begin{equation}
    \dot M_{\rm BH}=\frac{\eta_{\rm acc}M_\star}{\mathcal C t_{\rm fb}}\times\begin{cases}
        \left(\frac{t}{t_{\rm fb}}\right)^{-\alpha},&~t<t_{\rm fb}\\
        \left(\frac{t}{t_{\rm fb}}\right)^{-5/3},&~t>t_{\rm fb}
    \end{cases}
    \label{eq:acc_rate}
\end{equation}
where $0\le\alpha<1$ is the early-time accretion index\footnote{\cite{2014ApJ...784...87S} pointed out that a slow-decaying accretion rate is possible due to disk internal kinematic viscosity, depending on the type of polytrope stars.}, and $\mathcal C = 3+2/(1-\alpha)$ is the normalization coefficient. We then parameterize the power reprocessed to the relativistic jet from the mass accretion using the energy conversion efficiency $\eta_j$, 
\begin{equation}
    L_j=\eta_j\dot M_{\rm BH}c^2.
    \label{eq:Lj}
\end{equation}
The accretion and jet power efficiencies, $\eta_{\rm acc}$ and $\eta_{j}$, remain uncertain as they depend on the dynamics of mass fallback, disk formation, {and the magnetic flux accumulation} \citep[e.g.,][]{2020ApJ...902..108M}. Meanwhile, $\eta_j$ may depend on the accretion rate and the SMBH spin \citep[see, e.g.,][]{2017ApJ...849...47L}; for simplicity, we assume it to be constant. Noting that these two parameters are degenerate with the stellar mass in terms of data fitting, we define the total jet energy as $\mathcal{E}_{j} = \int L_j dt = \eta_{j} \eta_{\rm acc} M_{\star} c^2 / 2$, which, instead of $\eta_j$ and $\eta_{\rm acc}$, will be treated as a free parameter.

\subsection{Jet dynamics}

It is useful to define a generic density profile of the external medium through which the relativistic jets propagate and decelerate. The interpretation of radio observations using the forward shock model demonstrates that the density profile as a function of the distance to the SMBH ($R$), $n \propto R^{-k}$ with $1.5 < k < 2.0$, is favored \citep[e.g.,][]{2023MNRAS.522.4028M,2024ApJ...965...39Y,2024ApJ...963...66Z}, \cy{while the analysis
of radio emissions in non-jetted TDEs suggests a little steeper profile $k=2.5$ \citep[e.g.,][]{2020SSRv..216...81A}}. Fitting the X-ray spectra and light curves of AT 2022cmc indicates that a fast jet with a Lorentz factor of $\Gamma = {\rm few} \times 10$ is typically needed \citep{2024arXiv240611513Y}, implying that the jet could penetrate the circumnuclear medium (CNM) and reach the interstellar medium (ISM). Therefore, we follow \cite{2024arXiv240611513Y} and connect the CNM to the ISM, which yields
\begin{equation}
    n_{\rm ext}=
    \begin{cases}
        n_{\rm ISM}\left(\frac{R}{R_{\rm cnm}}\right)^{-k},&~R<R_{\rm cnm}\\
        n_{\rm ISM}, & ~ R>R_{\rm cnm}.
    \end{cases}
\end{equation}
One possible source for the CNM within $R_{\rm cnm}$ is the wind emanating from pre-existing disks, which predicts the boundary $R_{\rm cnm}\sim10^{18}~\rm cm$ before merging into the ISM \citep{2020PhRvD.102h3013Y,2021ApJ...911L..15Y}. \cy{Moreover, the wind density profile can be smoothly connected to ISM at the Bondi radius \citep{2024ApJ...971...49M}, which predicts the similar $R_{\rm cnm}$ of $\sim10^{18}$ cm for $M_{\rm BH}\sim10^7~M_\odot$}. {The density profile within $R_{\rm cnm}$ primarily impacts jet evolution in the very early stage and does not significantly influence the results after $T_{\rm obs}\sim$ few days.} In subsequent calculations, we adopt the fiducial values of $k=1.8$ \cy{as in \cite{2023MNRAS.522.4028M}} and $R_{\rm cnm}=10^{18}~\rm cm.$ 

We follow the methodology for blastwave dynamics, as illustrated in GRB afterglow modeling \citep{2013MNRAS.433.2107N,zhang2018physics,2023arXiv231113671Z,2024arXiv240413326S} and the structured jet modeling of AT 2022cmc \citep{2024arXiv240611513Y}, to model the jet dynamics. The jet has a top-hat structure and points towards the observer (on-axis). Specifically, we examine how the jet Lorentz factor and radius depend on time, given the initial Lorentz factor $\Gamma_0$, the jet opening angle $\theta_j$, and $n_{\rm ext}$. In this picture, as the jet penetrates deeply into the ambient gaseous environment, it sweeps up material, leading to the formation of the forward shock (FS), which accumulates and accelerates the upstream external medium to a Lorentz factor $\Gamma < \Gamma_0$. Meanwhile, a reverse shock (RS) decelerates the unshocked ejecta from $\Gamma_0$ to $\Gamma$ within the jet.

For jetted TDEs, we consider a continuously powered jet, in the sense that the jet energy $\mathcal E_j$ and ejecta mass are persistently injected from the central engine. 
A comprehensive treatment of jet evolution incorporating the reverse shock and continuous injections is presented in Appendix A of \cite{2024arXiv240611513Y}. Here, we numerically solve the coupled differential equations derived in \cite{2024arXiv240611513Y}. The results are consistent with the analytical solutions obtained from $\int L_jdt\propto \Gamma^2m_pn_{\rm ext}c^2R_j^3$ and $R_j\approx 2\Gamma^2ct$,
\begin{equation}
\Gamma(t)\propto\begin{cases}
    t^{-(2+\alpha)/8}, &~ t<t_{\rm fb}\\
    t^{-3/8}, &~ t> t_{\rm fb},
\end{cases}
\end{equation}
where $R_j$ is the radius of the jet.
In the following sections, the jet evolutions, specifically, $\Gamma(t)$ and $R_j(t)$, will be used to compute the time-dependent FS and RS emissions.

\subsection{Multi-wavelength emission from RS regions}

We focus on the RS scenario, as \cite{2024arXiv240611513Y} has revealed that the fast-decaying X-ray afterglow of AT 2022cmc could be attributed to emissions generated in RS regions. Given the isotropic equivalent jet luminosity, $L_{j,\rm iso}=L_j/(\theta_j^2/2)$, jet radius $R_j$ and Lorentz factor $\Gamma$, we estimate the relative Lorentz factor between the RS upstream and downstream, $\Gamma_{\rm rel}\approx(\Gamma_0/\Gamma+\Gamma/\Gamma_0)/2$, and the upstream particle number density \cy{$n_{0}'=L_{j,\rm iso}/(4\pi R_j^2\Gamma_0^2m_pc^3)$} in the jet comoving frame. The downstream magnetic field strength can be parameterized as $B_{\rm rs}=\sqrt{32\pi\epsilon_B\Gamma_{\rm rel}(\Gamma_{\rm rel}-1)n_{0}'m_pc^2}$, where $\epsilon_B$ represents the fraction of the internal energy density that is converted to magnetic field energy density. 

We consider a power-law injection rate, $\dot Q_e\propto\gamma_e^{-s}$, to describe the distribution of non-thermal electrons accelerated by relativistic shocks. In this expression, $\gamma_e$ is the electron Lorentz factor and $s\ge2$ is the spectral index. To normalize the injection rate, we introduce the number fraction $(f_e)$ of downstream electrons that are accelerated by RS and the energy fraction ($\epsilon_e$) of internal energy that is converted to non-thermal electrons. This allows us to infer the minimum Lorentz factor of injected electrons, \cy{$\gamma_{e,\rm m}=(\Gamma_{\rm rel}-1)(s-2)/(s-1)(\epsilon_e/f_e)(m_p/m_e)$ for $s>2$}, and normalize $\dot Q_e$. Using $B_{\rm rs}$, $\dot Q_{e}$, $R_j(t)$, and $\Gamma(t)$, we compute the time-dependent synchrotron and inverse Compton emissions in the synchrotron self-Compton (SSC) framework, and convert quantities in jet comoving frame to the observer's frame by integrating over the equal-arrival-time surfaces. We also account for the effect of jet break when the jet Lorentz factor drops below $1/\theta_j$ by applying $f_{\rm br}=1/[1+(\theta_j\Gamma)^{-2}]$ to the light curves\footnote{\cy{Here, we neglect the lateral spreading of the jet, known as the sideways expansion, for simplicity.}}. This steepens the post-break light curves after $t_{\rm br}$, defined by $\Gamma(t_{\rm br})=1/\theta_j$, with the factor $f_{\rm br}\propto \Gamma^2$ for $t>t_{\rm br}$. A detailed description of the numerical modeling of RS synchrotron and SSC emissions can be found in Appendix \ref{app:ssc}.

\begin{deluxetable}{c|c|c|c|chlDlc}

\tablenum{1}
\tablecaption{Physical parameters\label{tab:params} ($f_e=1.5\times10^{-3}$, $s=2.3$ and $\epsilon_e=0.2$ are fixed for all TDEs) and results} 
\tablewidth{0pt}
\tablehead{
{\bf TDEs $^{(a)}$} & {\bf AT 2022cmc}& {\bf J1644} & {\bf J2058}  &{\bf J1112} 
}
\startdata
$z$ & 1.19 & 0.35 & 1.19 & 0.89\\
 $M_{\rm BH}~[M_\odot]$ & $10^7$ & $10^6$ & $10^6$ & $2\times10^6$\\
  \hline
\multicolumn{3}{l}{\bf Model parameters}\\
   \hline
   $\alpha$ & 0.80  & 0.65 & 0.85 & 0.70\\
$\mathcal E_j~[10^{52}~\rm erg]$ & $5.4$ & 3.5 & 2.9 &  6.3\\
$n_{\rm ISM}~\rm [cm^{-3}]$ & 10 & 6.0 & 1.0 & 10\\
$\theta_j$ & 0.15 & 0.1 & 0.1 & 0.1\\
$\Gamma_{0}$ & 30 & 25 & 42 & 35\\
$\epsilon_{B}$ & 0.10 & 0.15 & 0.20 & 0.15\\
   \hline
\multicolumn{3}{l}{\bf Results}\\
   \hline
   $M_\star^{(b)}~[M_\odot]$ & 3.0  & 1.9 & 1.6 & 3.5\\
$T_{\rm fb}~[\rm d]$ & $77$ & 28 & 27 &  21\\
$T_{\rm br}~[\rm d]$ & $79$ & 212 & 76 &  37\\
$T_{\rm ce}^{(c)}~[\rm d]$ & $227$ & 352 & 331 &  470\\
\enddata
\footnotesize{$^{(a)}$ Data sources: AT 2022cmc \citep{2022Natur.612..430A,2023NatAs...7...88P}, Sw J1644 \citep{burrows2011relativistic}, Sw J2058 \citep{2015ApJ...805...68P}, and Sw J1112 \citep{2015MNRAS.452.4297B}.

$^{(b)}$ The fiducial value $\eta_j\eta_{\rm acc}=0.02$ is used to infer $M_\star$ from $\mathcal E_j$.

$^{(c)}$ {The fiducial parameter $\eta_{\rm rad}\eta_{\rm acc}=0.05$ is used to calculate the cessation time $T_{\rm ce}=(1+z)t_{\rm ce}$ of the central engine. Please refer to the main text for the definitions of the fallback time $T_{\rm fb}$, the jet break time $T_{\rm br}$, and $T_{\rm ce}$.}}

\end{deluxetable}

\section{Reverse shock Scenario for four jetted TDE\lowercase{s}}\label{sec:results}
\begin{figure*}\centering
    \includegraphics[width=0.49\textwidth]{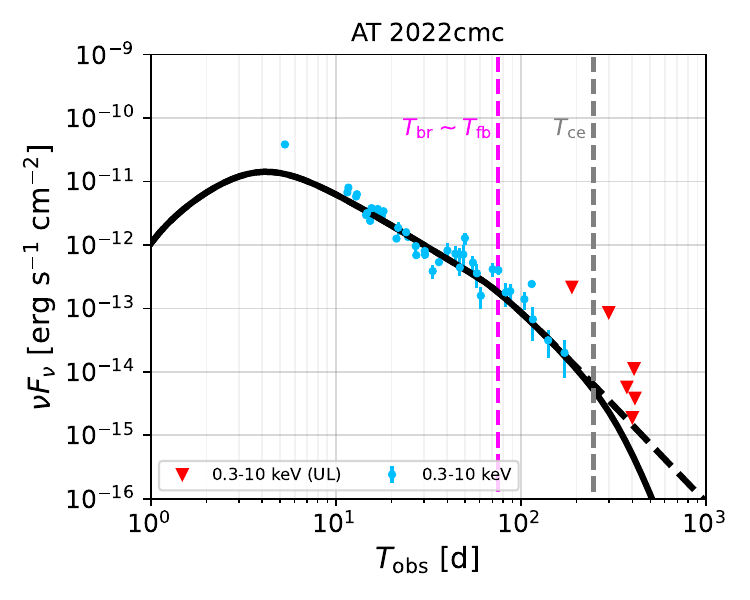}
    \includegraphics[width=0.49\textwidth]{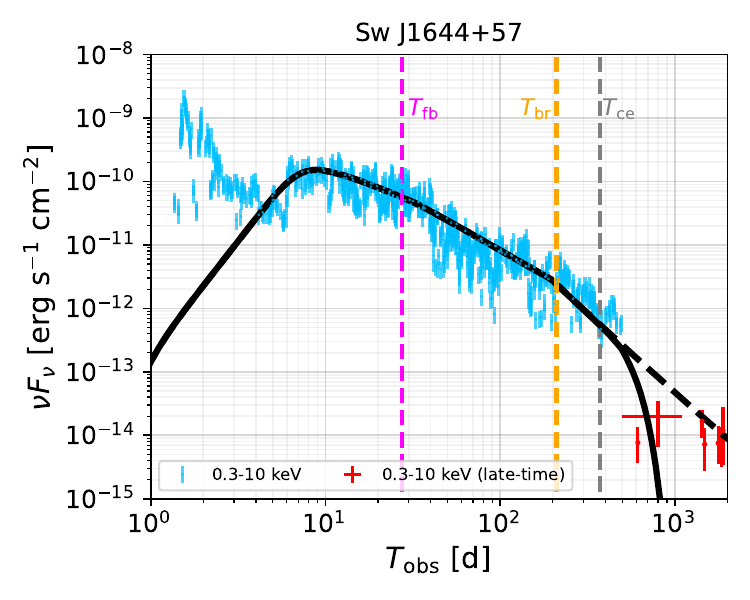}
    \includegraphics[width=0.49\textwidth]{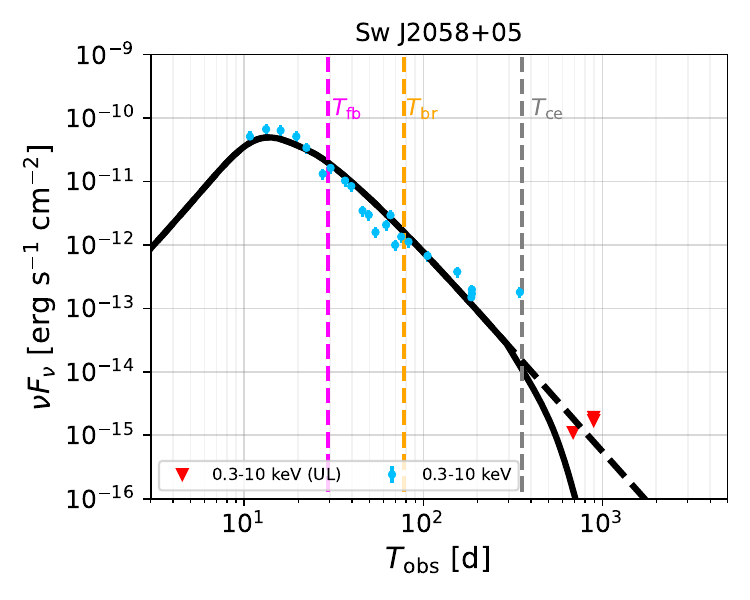}      
         \includegraphics[width=0.49\textwidth]{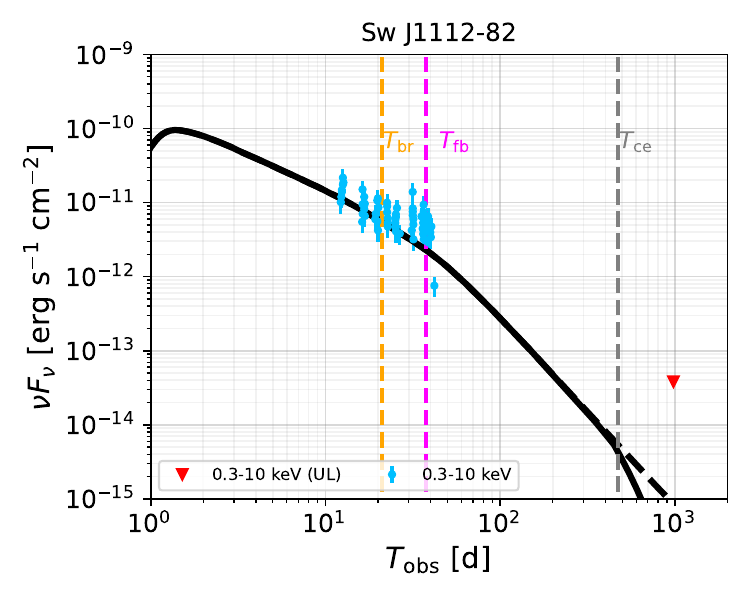}
         \caption{Swift 0.3 - 10 keV X-ray light curve fitting for four jetted TDEs. The early-stage and late-time data points (or upper limits) are shown as blue and red points (triangles). The dashed and solid black curves respectively depict the fitted X-ray light curves before and after accounting for the ceased central engine. The vertical dashed lines indicate the characteristic times obtained from the fitting, which are summarized in Table \ref{tab:params}. Data sources: AT 2022cmc \citep{2022Natur.612..430A,2023NatAs...7...88P, 2024ApJ...965...39Y, 2024arXiv240410036E}, Sw J1644 \citep{burrows2011relativistic,2013ApJ...767..152Z,2016ApJ...817..103M,2018ApJ...854...86E}, Sw J2058 \citep{2015ApJ...805...68P}, and Sw J1112 \citep{2015MNRAS.452.4297B}.}
                  \label{fig:X_LCs}

\end{figure*}

\subsection{X-ray data interpretations}\label{subsec:results_X}

Table \ref{tab:params} summarizes the observational parameters, such as the redshifts and the SMBH masses\footnote{{The SMBH masses are poorly constrained. These fiducial values are chosen to satisfy the X-ray variability constraints \citep[e.g.,][]{2024arXiv240410036E} and to ensure that the cessation times (defined later) are consistent with observations.}}, for AT 2022cmc, Sw J1644 (J1644), Sw J2058 (J2058) and Sw J1112 (J1112). Noting that the masses of disrupted stars are not efficiently identified and can degenerate with the energy conversion efficiencies of accretion activities and jets, we treat the total jet energy $\mathcal E_j=\int L_jdt=\eta_j\eta_{\rm acc}M_\star c^2/2$ as a free parameter to infer the time-dependent jet power using Eqs. (\ref{eq:acc_rate}) and (\ref{eq:Lj}). Additionally, we select the accretion rate index $\alpha$ to model the X-ray light curves before the mass fall back time $T_{\rm fb}=(1+z)t_{\rm fb}$ and choose $n_{\rm ISM}$, $\theta_j$, and $\Gamma_0$ as fitting parameters {to ensure that $T_{\rm fb}$ and jet break time $T_{\rm br}=(1+z)t_{\rm br}$ are consistent with the observed X-ray light curves (see the magenta and orange lines in Fig. \ref{fig:X_LCs})}. For multi-wavelength emissions from RS regions, we fix $f_e=1.5\times10^{-3}$, $s=2.3$ and $\epsilon_e=0.2$ to reduce the number of free parameters based on the interpretations of the AT 2022cmc X-ray data \citep{2024arXiv240611513Y}\footnote{These values are also consistent with values suggested by the RS model for early GRB afterglow emission \citep{Genet:2007nb}}, while we allow $\epsilon_B$ to vary freely for X-ray spectral fittings, as it significantly impacts the cooling frequencies in the synchrotron spectra.

Although the RS scenarios have been shown to describe the X-ray afterglows of jetted TDEs, and FS emissions are subdominant in X-ray ranges, we include here the contribution from FS regions for completeness. Instead of introducing new free parameters, we follow the treatment in \cite{2024arXiv240611513Y} and apply the FS parameters obtained from the AT 2022cmc radio data fitting to all four jetted TDEs. 

Applying the model described in Sec. \ref{sec:model} to X-ray light curves and spectra in the energy range of 0.3-10 keV for AT 2022cmc, Sw J1644, Sw J2058 and Sw J1112, we obtain the fitting parameters, as shown in Table \ref{tab:params}, where the characteristic times $T_{\rm fb}$ and $T_{\rm br}$, are also shown. Fig. \ref{fig:X_LCs} illustrates the fit of X-ray light curves. 
The black dashed curves show the X-ray light curves predicted by RS scenario with jet break corrections. The numerical results are consistent with the analytical RS light curves at $E_X\sim1$ keV \citep{2024arXiv240611513Y}, 
\begin{equation}
    \nu F_\nu\propto
    \begin{cases}
        f_{\rm br} T_{\rm obs}^{-[5\alpha+\alpha(s-1)]/4},~& T_{\rm obs}<T_{\rm fb}\\
        f_{\rm br} T_{\rm obs}^{-(2s+25)/12},~& T_{\rm obs}>T_{\rm fb}.
    \end{cases}
    \label{eq:X_LC}
\end{equation}
From this figure, we observe that the RS model with jet break steepening can describe the X-ray afterglows of AT 2022cmc and Sw J1112, encompassing both the early-stage observations and the late-time upper limits. However, for Sw J1644 and Sw J2058, the late-time X-ray emissions from the RS model exceed the upper limits (see the corresponding dashed black curves), indicating that a sharper decay is required.

\begin{figure*}[htp]
    \centering\includegraphics[width=0.49\textwidth]{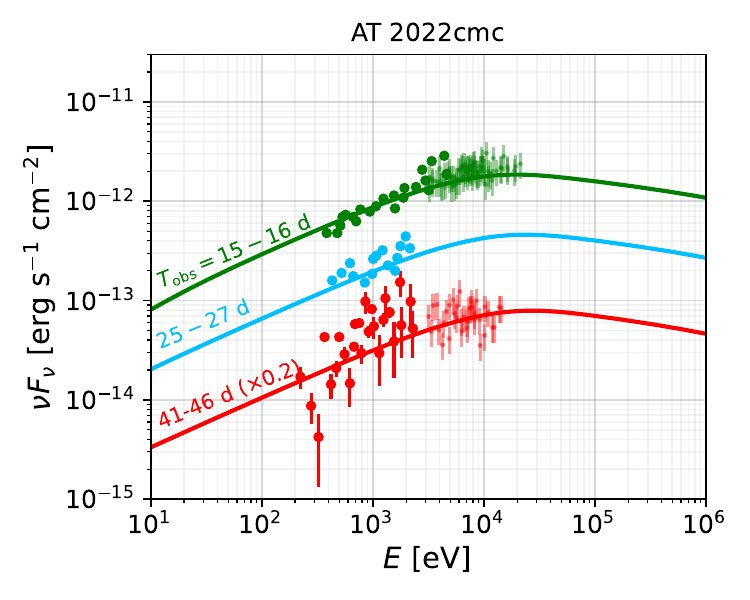}
\includegraphics[width=0.49\textwidth]{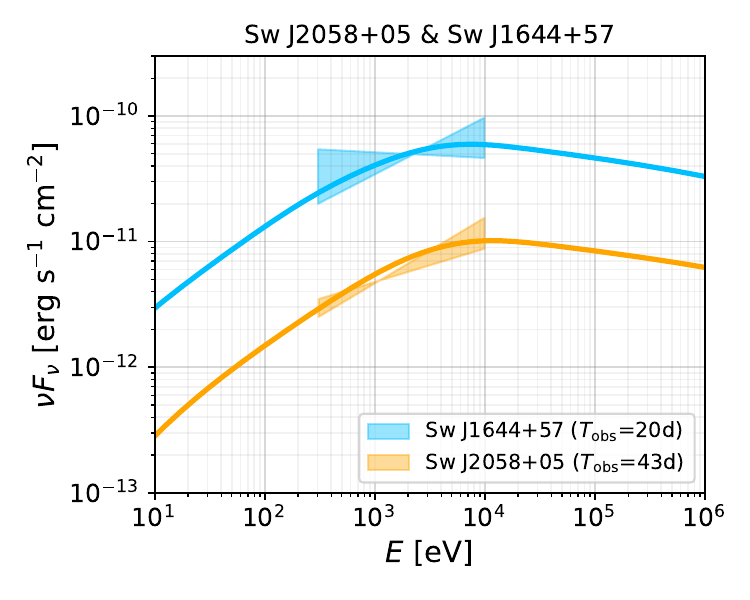}
\caption{X-ray spectra fitting for AT 2022cm (left panel), J1644 and J2058 (right panel) at different times. The data sources are the same as those of Fig. \ref{fig:X_LCs}.}
\label{fig:Xspec}
\end{figure*}

To address this issue, we introduce the ceased central engine mechanism by turning off the jet power injection $L_j$ after $T_{\rm ce}=(1+z)t_{\rm ce}$, {as the accretion rate becomes sub-Eddington \citep{2013ApJ...767..152Z,2014MNRAS.437.2744T}}. The cessation time $t_{\rm ce}$ is determined by $\dot M_{\rm BH}(t_{\rm ce})=L_{\rm Edd}/(\eta_{\rm rad}c^2)$, where $L_{\rm Edd}\simeq1.3\times10^{45}~{\rm erg~s^{-1}}M_{\rm BH,7}$ is the Eddington luminosity and $\eta_{\rm rad}$ is the radiation efficiency. \cy{Explicitly, we write down
\begin{equation}\begin{split}
        t_{\rm ce}&=\left(\frac{\eta_{\rm acc}\eta_{\rm rad}M_\star c^2}{\mathcal C L_{\rm Edd}t_{\rm fb}}\right)^{3/5}t_{\rm fb}\simeq134{~\rm d}~M_{\star,0.7}^{14/25}M_{\rm BH,7}^{-2/5}
        \end{split}
\end{equation} for $\alpha=0.8$ and $\eta_{\rm acc}\eta_{\rm rad}=0.05$}.
Table 2 summarizes $T_{\rm ce}$ for each TDE, using $\eta_{\rm acc}\eta_{\rm rad}=0.05$ as the fiducial parameter. Physically, after $t_{\rm ce}$, the RS emission decays rapidly, as the ceased central engine suspends the electron injection once the shock crossing is complete. The solid black curves in Fig. \ref{fig:X_LCs} illustrate the light curves after shutting down the power injection following $T_{\rm ce}$, while the gray dashed lines correspond to the cessation times for each TDE. We conclude that the RS model, incorporating jet break steepening and the ceased central engine mechanism, effectively explains the rapid decay observed in the late-stage X-ray light curves, particularly for Sw J2058 and Sw J1644. In both cases, the FS contributions are subdominant as pointed out by \cite{2024arXiv240611513Y} for AT 2022cmc.

\begin{figure*}[htp]
    \centering
    \includegraphics[width=0.49\textwidth]{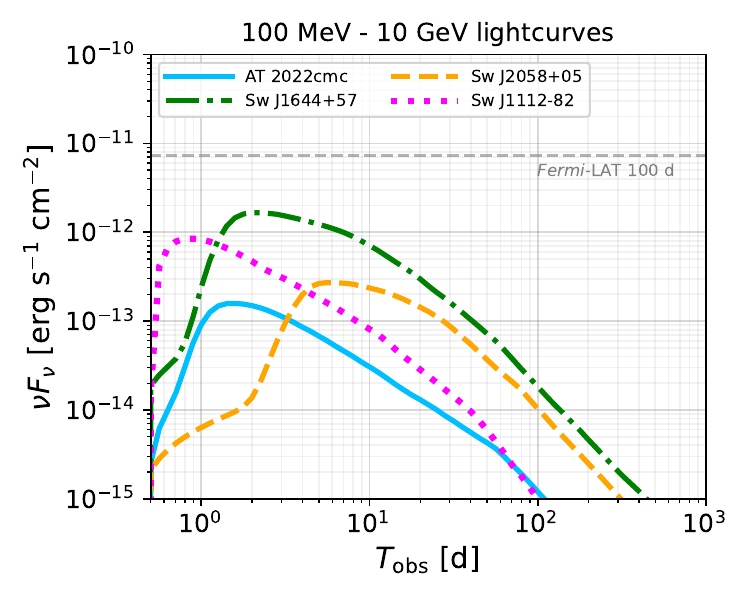}
    \includegraphics[width=0.49\textwidth]{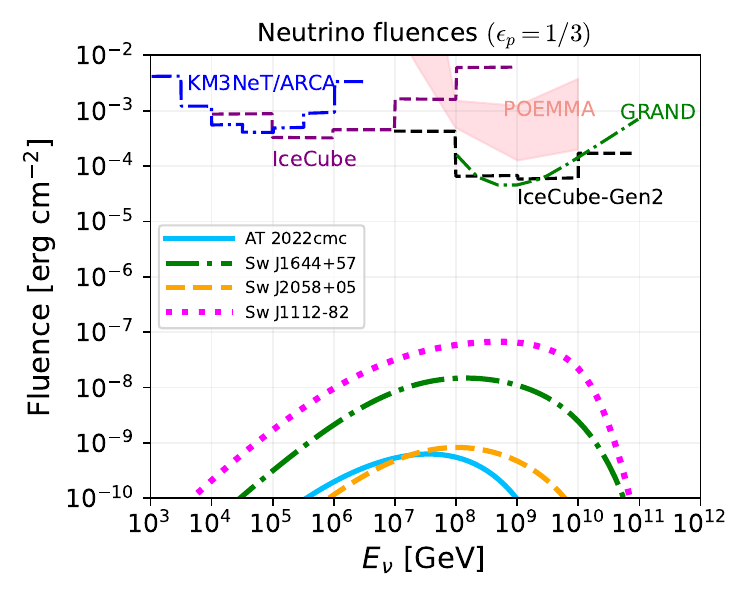}
    \caption{\emph{Left panel:} Model-predicted 100 MeV - 10 GeV $\gamma$-ray light curves for each TDE. The horizontal dashed line indicates the \emph{Fermi}-LAT sensitivity for 100 d observations. \emph{Right panel}: Expected single-flavor neutrino fluences originated from the external FS and RS regions. Optimistic proton acceleration efficiency of $\epsilon_p=1/3$ is used. In both calculations, the parameters obtained from X-ray data fits are applied. See the main text for a detailed description of the sensitivity curves.}
    \label{fig:gamma_ray}
\end{figure*}

In the early stage when the jet is still in the coasting phase, the X-ray fluxes increase with time as the RS begins to cross the ejecta. This feature explains the peaks observed in the light curves of Sw J1644 and Sw J2058 at $T_{\rm obs}\sim10$ d. \cy{The X-ray observations of AT 2022cmc at $T_{\rm obs}\sim4$ d and the fluctuations seen in Sw J1644's light curve at earlier times, e.g., $T_{\rm obs}\lesssim4$ d, may be attributed to internal energy dissipations occurring close to the SMBH \citep[e.g.,][]{2013ApJ...762...98L}, analogous to the prompt phase of GRBs.}
 
{We focus on spectral fitting in the X-ray range, e.g., 0.3-10 keV, as there are no $\gamma$-ray observations for these four jetted TDEs, and the radio observations are typically attributed to forward shock emissions from a slower jet \citep[e.g.,][for AT 2022cmc]{2023MNRAS.522.4028M, 2024ApJ...963...66Z, 2024arXiv240611513Y}.} We present in Fig. \ref{fig:Xspec} the fit to X-ray spectra for AT 2022cmc (left panel), Sw J2058 and Sw J1664 (right panel). The results indicate that the fast cooling synchrotron spectra can reproduce the X-ray spectra in the energy range 0.3-10 keV.

\subsection{$\gamma$-ray and neutrino detectabilities}\label{subsec:results_MM}

The accelerated non-thermal electrons could produce $\gamma$-rays in energy ranges of the \emph{Fermi} Large Area Telescope (LAT), e.g., 100 MeV to 10 GeV. It is useful to discuss the $\gamma$-ray detectabilities. The left panel of Fig. \ref{fig:gamma_ray} illustrates the time-dependent $\gamma$-ray flux for each jetted TDE. Similar to the X-ray light curves, the $\gamma$-ray light curves exhibit fast-decaying signatures with the peak flux level maintaining over tenths of days. For reference, the 100 d \emph{Fermi}-LAT sensitivity line \citep{2021ApJS..256...12A} is also shown as the horizontal gray line. This figure indicates that it is challenging to detect the fast-decaying $\gamma$-ray emissions by \emph{Fermi}-LAT, which explains the non-detection of these sources in $\gamma$ rays. In the very-high-energy $\gamma$-ray regime, e.g., $\gtrsim$100 GeV - 1 TeV, the fluxes are suppressed by Klein-Nishina effect of inverse Compton radiation and $\gamma\gamma$ attenuation with extragalactic background lights (EBLs), making these jetted TDEs increasingly challenging to be detected unless the source is nearby and/or additional mechanisms such as external inverse Compton emission are introduced.

Recent identification of TDEs and TDE candidates with potential neutrino correlations, e.g., AT 2019dsg \citep{2021NatAs...5..510S}, AT 2019fdr \citep{2022PhRvL.128v1101R}, AT 2019aalc \citep{2021arXiv211109391V}, AT 2021lwx \citep{2024ApJ...969..136Y}, ATLAS17jrp \citep{Li:2024qcp} and two obscured candidates \citep{2023ApJ...953L..12J}, have revealed that TDEs could be promising neutrino emitters. 
Here, {motivated by these observations, we consider neutrino emission from on-axis jetted TDEs.} We use the parameters obtained from X-ray data fittings to compute the cumulative single-flavor neutrino fluences from external relativistic shocks of jetted TDEs, which are shown in the right panel of Fig. \ref{fig:gamma_ray}. {Neutrino mixinig in vacuum is assumed.}
In this calculation, we consider an optimistic case where 1/3 of the jet kinetic power is converted to the accelerated protons (e.g., $\epsilon_p=L_p/L_j=1/3$) and treat the {synchrotron} emissions as the target photon field for $p\gamma$ interactions. For simplicity, we assume a power-law proton injection spectrum $\dot Q_p\propto E_p^{-2}\exp(-E_p/E_{p,\rm max})$ and determine the maximum energy $E_{p,\rm max}$ by equating the acceleration time to the cooling time. 

In the right panel of Fig. \ref{fig:gamma_ray}, the sensitivity curves of {IceCube (declination $\delta=-23^\circ$), IceCube-Gen2 \citep[$\delta=0^\circ$,][]{IceCube-Gen2:2021rkf}, GRAND \citep[zenith angle $\theta=90^\circ$,][]{2020SCPMA..6319501A}, KM3NeT/ARCA230 \citep[$\delta=-73^\circ$,][]{KM3NeT:2024uhg} and POEMMA \citep[90\% unified confidence level,][]{2020PhRvD.102l3013V}} are also shown. We find that the neutrino fluences from these jetted TDEs are at least two orders of magnitude lower than the IceCube-Gen2 sensitivity. 
The major reason is that the jetted TDEs are fast-fading transients compared to persistent AGNs, and the $p\gamma$ interaction efficiencies in the jets of radii $R_{j}\gtrsim10^{18}$ cm are low unless a substantial number of external target photons are introduced. Compact jet, internal energy dissipations, external target photons, and/or dense external media \citep[e.g.,][]{2024arXiv240313902K}, would be needed to make jetted TDEs promising neutrino emitters.

{We also note that our model would not explain neutrino events from AT 2019dsg, AT 2019fdr, AT 2019aalc, and AT 2021lwx because the fleunces are too small. In addition, these TDEs are not on-axis jetted TDEs and the jet energy is constrained by afterglow observations~\citep{2020ApJ...902..108M}, although the constraints are relaxed hidden jets with time delays~\citep{2024MNRAS.534.1528M}.}

\section{Discussion}\label{sec:discussion}
We have presented the reverse shock model incorporating continuously powered jets to explain the X-ray afterglows of four jetted TDEs: AT 2022cmc, Sw J1644, Sw J2058, and Sw J1112. 
Concerning the number of free parameters, we observe that by fixing $f_e=1.5\times10^{-3}$, $s=2.3$, $\epsilon_e=0.2$, and $k=1.8$, the spectral and light curve fitting reduces the degeneracy of the parameters. For instance, $\alpha$ determines the light curve slopes before $T_{\rm fb}$, while $\mathcal E_j$ normalizes the X-ray luminosities. The parameters $n_{\rm ISM}$ and $\Gamma_0$ jointly define the jet deceleration time and the light curve peaks, and $\theta_j$ controls the jet break time $T_{\rm br}$, whereas the cooling frequencies of X-ray spectra depend strongly on $\epsilon_B$. Moreover, by adopting the typical value $\eta_{\rm acc}\eta_{\rm j}\sim0.02$, we infer that the masses of the disrupted stars are distributed in the range $1.6-3.5~M_\odot$ (see Table \ref{tab:params}), which is consistent with the limits estimated from optical/radio observations.


In addition to the rapid decays, another significant feature of the observed X-ray light curves of {the 4 jetted TDEs, especially Sw J1644, is their rapid variability of $\sim{\rm a~few\times}100-1000$ s}. The variability time scale of homogeneous reverse shock downstreams is typically close to the observation time, e.g.,
$T_{\rm rs,var}\sim (1+z)R_j/(\Gamma^2c)\sim T_{\rm obs}$. Since the jet is continuously powered by the SMBH, the observed variability time is modulated by the light crossing time of the central engine \citep{2005ApJ...631..429I,2012Sci...337..949R}. Moreover, the stochastic magnetic dissipation \citep{2011ApJ...726...90Z} and small scale plasma fluctuations may also cause short-term variabilities. On the other hand, the X-ray light curves also exhibit some quasi-periodic ($1-10$ days) variation, with dips that cannot be simply attributed to external shock evolution. These short-term structures may be caused by jet precessions induced by misaligned accretion disks \citep{2012PhRvL.108f1302S,2012MNRAS.422.1625S,2013ApJ...762...98L,2024MNRAS.528.2568C}; however a quantitative modeling of these effects is beyond the scope of this work. 

Besides the RS emission, contributions from the FS region are also included in our calculations and are found to be negligible in X-ray bands using the same parameters for radio data fitting, which disfavors the external FS as the origin of jetted TDE X-ray emission. 
This also supports the two-component jet model, suggesting that the FS in slow ($\Gamma_0\sim1-10$) jets and the RS in fast ($\Gamma_0>10$) jets explain the radio and X-ray observations \citep{2024arXiv240611513Y}, respectively.

\section{Conclusions}\label{sec:conclusion}

We modeled the jet evolution within an external density medium by connecting the CNM profile, e.g., $n_{\rm cnm}\propto R^{-k}$, to the ISM, and computed the time-dependent synchrotron and SSC emission from reverse shock regions. 
We concluded that the external reverse shock model for the jet may provide a generic and self-consistent framework to explain the baselines of the X-ray afterglows of jetted TDEs: Sw J1644, Sw J2058, Sw J1112, and AT 2022cmc. Remarkably, the jet break, together with a ceased central engine when the accretion rate falls below the Eddington accretion rate, could explain the sharp declines in the late-stage X-ray light curves, especially for Sw J1644 and Sw J2058. 

Our work also predicts the peak $\gamma$-ray flux to be $\lesssim2\times10^{-12}~\rm erg~s^{-1}~cm^{-2}$ (see the left panel of Fig. \ref{fig:gamma_ray}), which is lower than the \emph{Fermi}-LAT 100 d sensitivity and explains the non-detection of $\gamma$-rays from jetted TDEs. Moreover, chosing an optimistic proton acceleration efficiency of $\epsilon_p=L_p/L_j=1/3$, the neutrino fluences from the four jetted TDEs are still more than two orders of magnitude lower than the IceCube-Gen2 sensitivity.  

Future work incorporating external reverse shocks, periodic jet precessions, and the contribution of internal energy dissipations would offer a more comprehensive description of the TDE X-ray afterglows, reproducing simultaneously the overall slopes, short-term fluctuations, and rapid variability. 


\acknowledgments

We would like to thank Brian Metzger for useful discussions, and Damiano Fiorillo for useful comments on this manuscript. 
The work of K.M. is supported by the NSF grant Nos.~AST-2108466, AST-2108467, and AST-2308021, and the JSPS KAKENHI grant No.~20H05852.

\software{NumPy \citep{NumPy_ref}, Matplotlib \citep{Matplotlib_ref}, pybind11 \citep{pybind11}, Eigen \citep{eigenweb}, AM$^3$ \citep{2023arXiv231213371K}}

\bibliography{ref.bib}
\bibliographystyle{aasjournal}

\appendix 
\section{Time-dependent Reverse shock Emissions}\label{app:ssc}
Given the minimum electron Lorentz factor, $\gamma_{e,\rm m}=(\Gamma_{\rm rel}-1)g(s)(\epsilon_e/f_e)(m_p/m_e)$, we normalize the electron injection rate $\dot Q_e$
via
\begin{equation}
    (4\pi R_j^2t_{\rm dyn}'c)\int \dot Q_e d\gamma_e=\frac{f_e L_{j,\rm iso}}{\Gamma_0m_pc^2},
    \label{eq:elec_norm}
\end{equation}
where $t_{\rm dyn}'=R_j/(\Gamma c)$ is the comoving dynamic time, $g(s)=(s-2)/(s-1)$ for $s>2$ and $g(s)\sim\mathcal O(0.1)$ for $s=2$.

We use the AM$^3$ software \citep{2023arXiv231213371K} to model the time-dependent synchrotron and inverse Compton emissions in the synchrotron self-Compton (SSC) framework by numerically solving the transport equations for electrons and photons in the comoving frame, specifically, for electrons,
\begin{equation}
    \frac{\partial n_e'}{\partial t'}=\dot Q_e-\frac{\partial}{\partial \gamma_e}(\dot\gamma_en_e')-\frac{n_e'}{t_{\rm dyn}'},
\end{equation}
where $n_e'$ is the electron number density (differential in Lorentz factor and volume), $t'$ is the time measured in comoving frame, and $\dot\gamma_e$ is the electron energy loss rate due to synchrotron radiation, inverse Compton scattering, and adiabatic cooling. In this calculation, we self-consistently determine the maximum electron Lorentz factor by equating the acceleration rate $t_{\rm acc}'^{-1}=eB_{\rm rs}/(\gamma_em_ec)$ to the cooling rate $t_{c}'^{-1}=|\dot\gamma_e|/\gamma_e$.

To obtain the observed photon spectra and light curves, we convert the comoving photon density spectra $n_\gamma'$ (in the units of $\rm cm^{-3}$) to the flux in the observer's frame by
integrating over the equal-arrival-time surfaces \citep[EATSs, e.g.,][]{Takahashi:2022xxa,2023arXiv231113671Z,2024arXiv240413326S}, 
\begin{equation}
\begin{split}
       F(E,T_{\rm obs})&=\frac{2\pi(1+z)}{d_L^2}\int_0^{\theta_j}d\theta \sin(\theta)R_j^2
       \\
       &\times\Delta R_jf_{\rm br}\frac{j'(\hat t,R_j,\varepsilon')}{\Gamma^2(1-\beta\mu)^2}\Big|_{\hat t=\frac{T_{\rm obs}}{1+z}+\frac{\mu R_j}{c}},
    \end{split}
    \label{eq:EATS}
\end{equation}
where $j'=n_\gamma'/(4\pi t'_{\rm dyn})$ is the emissivity per unit solid angle, $\hat t$ is the emission time,  $\mu=\cos\theta$, $\beta=\sqrt{1-1/\Gamma^2}$, $\beta_{\rm sh}=\sqrt{1-1/\Gamma_{\rm sh}^2}$ is the shock velocity with $\Gamma_{\rm sh}=\sqrt{2}\Gamma$. In this expression, $\varepsilon'=(1+z)E/\Gamma$ connects the energy in observer's frame ($E$) to energy in comoving frame ($\varepsilon'$), and $\Delta R_j={R_j}/{[12\Gamma^2(1-\mu\beta_{\rm sh})]}$ \citep{Takahashi:2022xxa} measures the radical thickness of the shocked region that contributes to the observed flux at $T_{\rm obs}$. The jet break correction factor $f_{\rm br}=1/[1+(\theta_j\Gamma)^{-2}]$ is also included.

\end{CJK*}
\end{document}